\documentclass{appolb}
\usepackage{epsfig}
\usepackage{5_here}
\usepackage{cite}

\begin{document}
\title{COMPARATIVE STUDIES OF LOW ENERGY $pp\eta$ AND $pp\eta^{\prime}$ SYSTEMS WITH COSY-11
DETECTOR%
\thanks{Presented at the Symposium on Meson Physics, Cracow, 01-04 October 2008.}%
}
\author{Pawe{\l}~Klaja, Rafa{\l}~Czy{\.z}ykiewicz and Pawe{\l}~Moskal\\
 on behalf of the COSY-11 collaboration
\address{Institute of Physics, Jagiellonian University, Cracow, Poland}
}
\maketitle
\begin{abstract}
We present a comparison of the two-body invariant mass distributions for the $pp \to pp\eta$ and 
$pp \to pp\eta'$ reactions, both measured at a nominal 
excess energy value of Q = 15.5 MeV.
For the $pp \to pp\eta$ reaction, in addition, the differential cross sections were
extracted for an excess energy of Q = 10 MeV. \\
The comparison of the results for the $\eta$ and $\eta^{\prime}$ meson production
rather
excludes the hypothesis that the enhancement observed in the invariant mass distributions
is due to the interaction of the meson and the proton.

\end{abstract}
\begin{figure}[H]
  \includegraphics[height=.33\textheight]{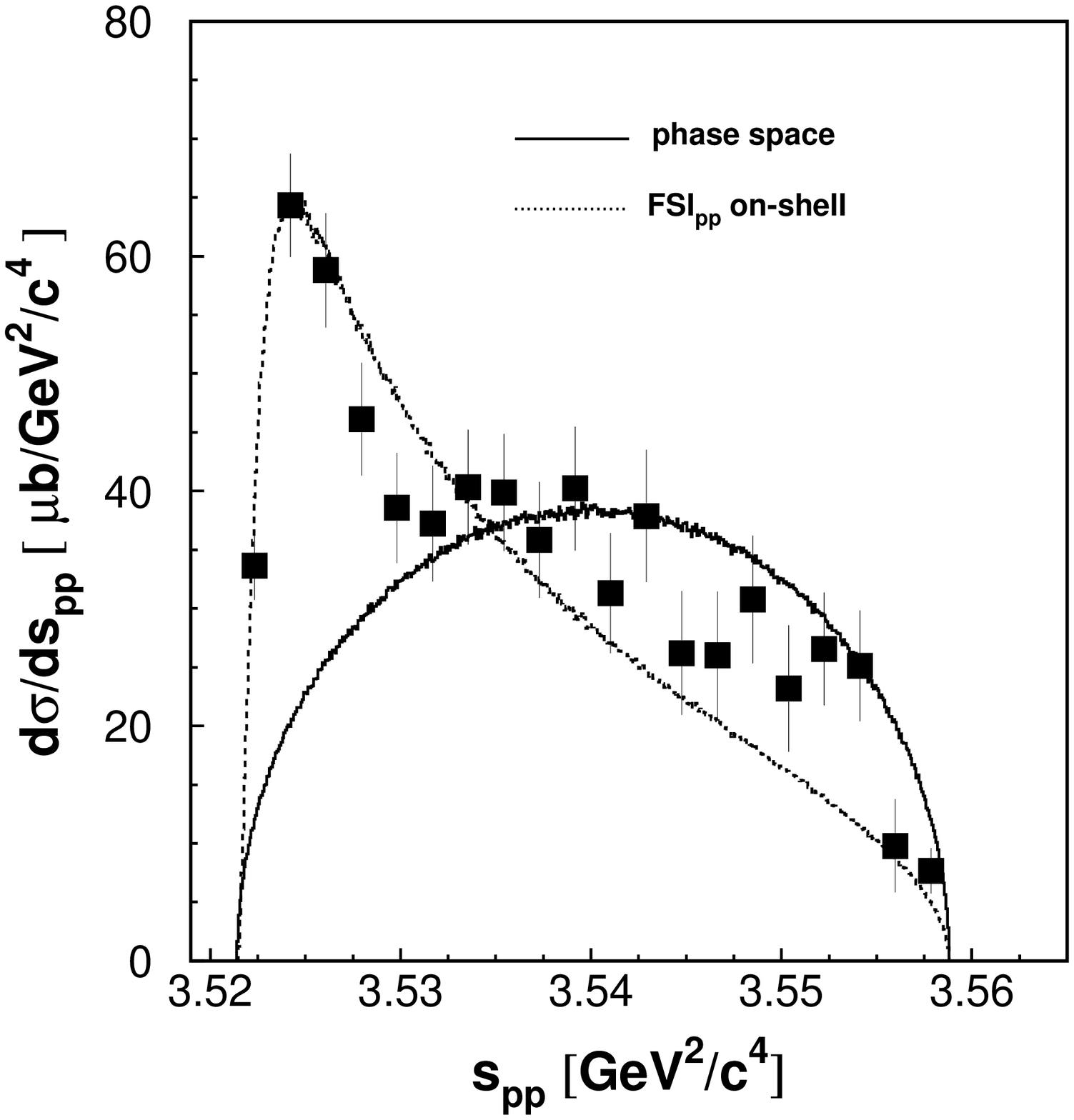}
  \includegraphics[height=.33\textheight]{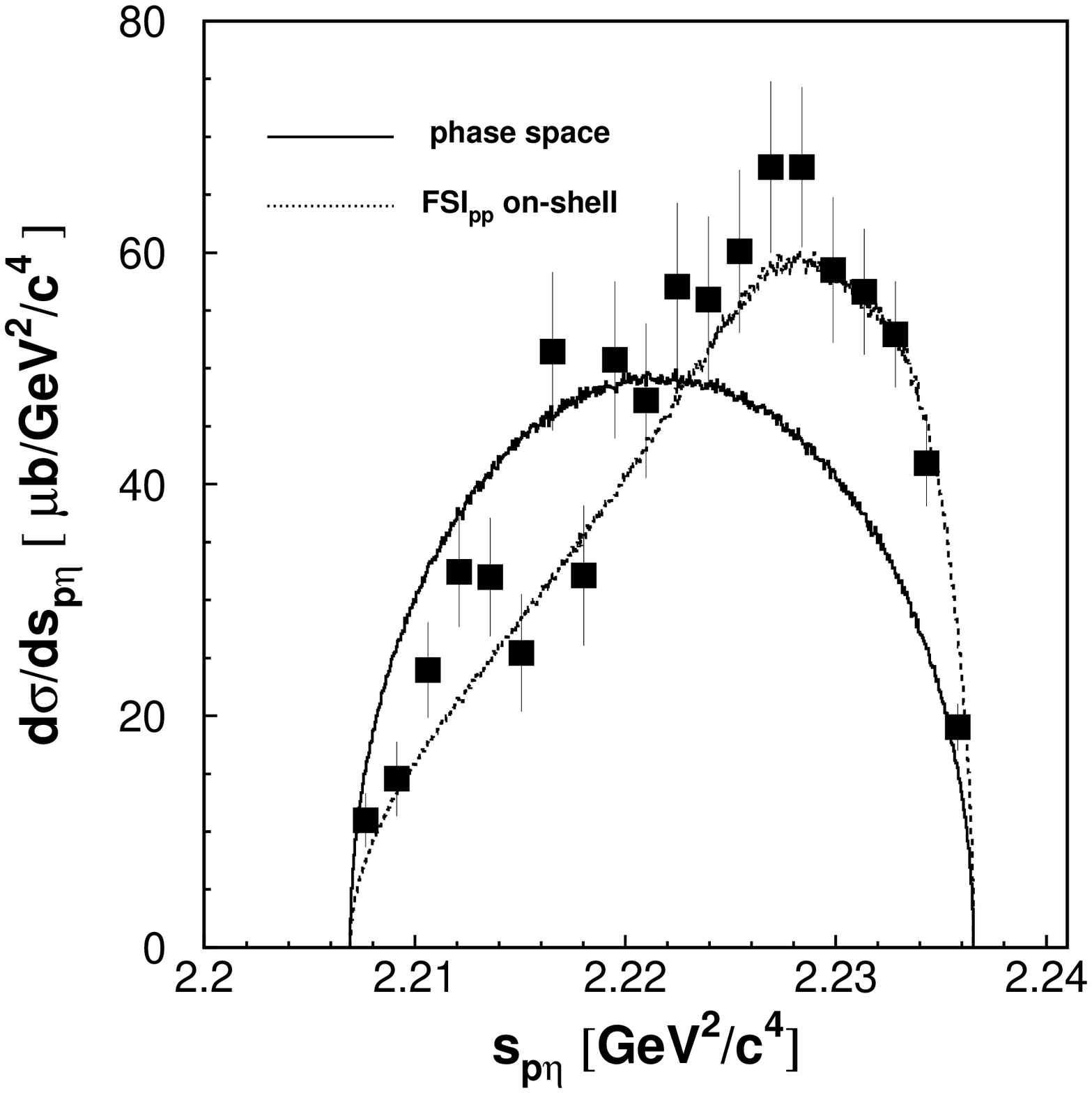}
  \caption{Distributions of the square of the proton-proton ($s_{pp}$) (left) and
  proton-$\eta$ ($s_{p\eta}$) (right) invariant masses determined experimentally for
  the $pp \to pp\eta$ reaction at Q = 10 MeV (full squares). The integrals of the phase space weighted by the
  square of the proton-proton on-shell scattering amplitude
(dotted lines), have been normalized arbitrarily at small values of $s_{pp}$.
The expectation
under the assumption of a homogeneously populated phase space are shown as thick solid lines.}
\label{czyzyk}
\end{figure}

\vspace{0.5cm}
The COSY-11 collaboration carried out 
experiments aming at the understanding of the near threshold meson production mechanisms, 
the meson--nucleon interaction and the meson structure. 
One specific part of the
COSY-11  physics program is devoted to the comparative study of
the interaction within the $pp\eta$ and $pp\eta'$ systems created near the kinematical threshold. \\
Near the threshold measurements of nucleon-nucleon collisions allow to study
particle production with a dominant contribution from one partial wave only \cite{review}.
Also, the interaction between particles in  near threshold collisions determines
strongly the dependence of the total cross section as a function of the centre-of-mass excess energy.
The experimentally determined excitation functions for the $pp \to pp\eta'$ \cite{prime, hibou} and
$pp \to pp\eta$ \cite{hibou, chiav, berg, smyrski, calen} reactions compared 
to the arbitrarily normalized phase-space integral reveals that proton-proton
FSI enhances the total cross section by more than one order of magnitude for low energies.
In the case of the $\eta'$ meson production the data are described
well assuming that the on-shell proton-proton amplitude exclusively determines the phase-space population.
In the case of the $\eta$ meson the pp-FSI is not sufficient for the description 
of the threshold enhancement of the excitation function.
These observations indicate that the proton-$\eta$ interaction is larger then the proton-$\eta'$ interaction 
and that the latter is too small to  manifest itself in
the excitation function within the presently achieved statistical uncertainty \cite{swave,habil}.
The interaction between particles depends on their relative momenta or equivalently on
the invariant masses of the two-particles subsystems. It should manifest itself as modification
of the phase-space abundance in kinematical regions where particles have small relative velocities.
Indeed, a qualitative phenomenological analysis of the determined
differential squared invariant proton-proton and proton-$\eta$ mass
distributions for the $pp \to pp\eta$ reaction measured by the COSY-11
collaboration at an excess energy of 15.5 MeV revealed an
enhancement of the population density at the kinematical region corresponding to small
proton-$\eta$ momenta~\cite{prc69}.\\ 
Also for the COSY-11 measurements performed at an excess energy of 4.5~MeV
a similar
enhancement has been observed~\cite{prc69}. 
In this contribution we present new results (see figure~\ref{czyzyk}) of  the squared invariant proton-proton and proton-$\eta$ mass
distribution determined at Q~=~10 MeV. The results are derived from the data analyses previously 
in view of the analysing power~\cite{prl, czyzyk_phd}.
The dashed lines in both panels of figure~\ref{czyzyk} depict the result of the
calculations where only the proton-proton interaction has been taken into account.
In those calculations the enhancement factor has been calculated as the square of the on-shell
proton-proton scattering amplitude, derived using the modified Cini-Fubini-Stanghelini formula
including the Wong-Noyes Coulomb corrections~\cite{swave}.\\
One can see that also at Q = 10 MeV, the discussed enhancement occurs to be too large
to be described by the on-shell inclusion of the proton-proton FSI \cite{swave}.\\
The observed enhancement could be explained by a significant role of the proton-$\eta$ interaction \cite{fix,fix2}
in the final state, or by an admixture of higher partial waves \cite{kanzo}, or by 
a possible energy dependence of the production amplitude \cite{deloff}. However, based on the spin-averaged $pp \to pp\eta$ 
data it is impossible to disentangle between the proposed hypothesis.\\
This 
motivated the measurement of the $pp \to pp\eta'$ reaction in order to
determine the distribution
of events over the phase space at an excess energy equal to 15.5~MeV, the same as for one of the measurements 
of the $pp \to pp\eta$ reaction \cite{prc69}.
The comparison of differential distributions of proton-proton and proton-meson invariant masses
for the $\eta$ and $\eta'$ production could help to judge between the postulated explanations
of the observed effect and in addition could allow for a quantitative estimation of the interaction between proton-$\eta$ and
proton-$\eta'$. 
\begin{figure}[H]
 \includegraphics[height=.33\textheight]{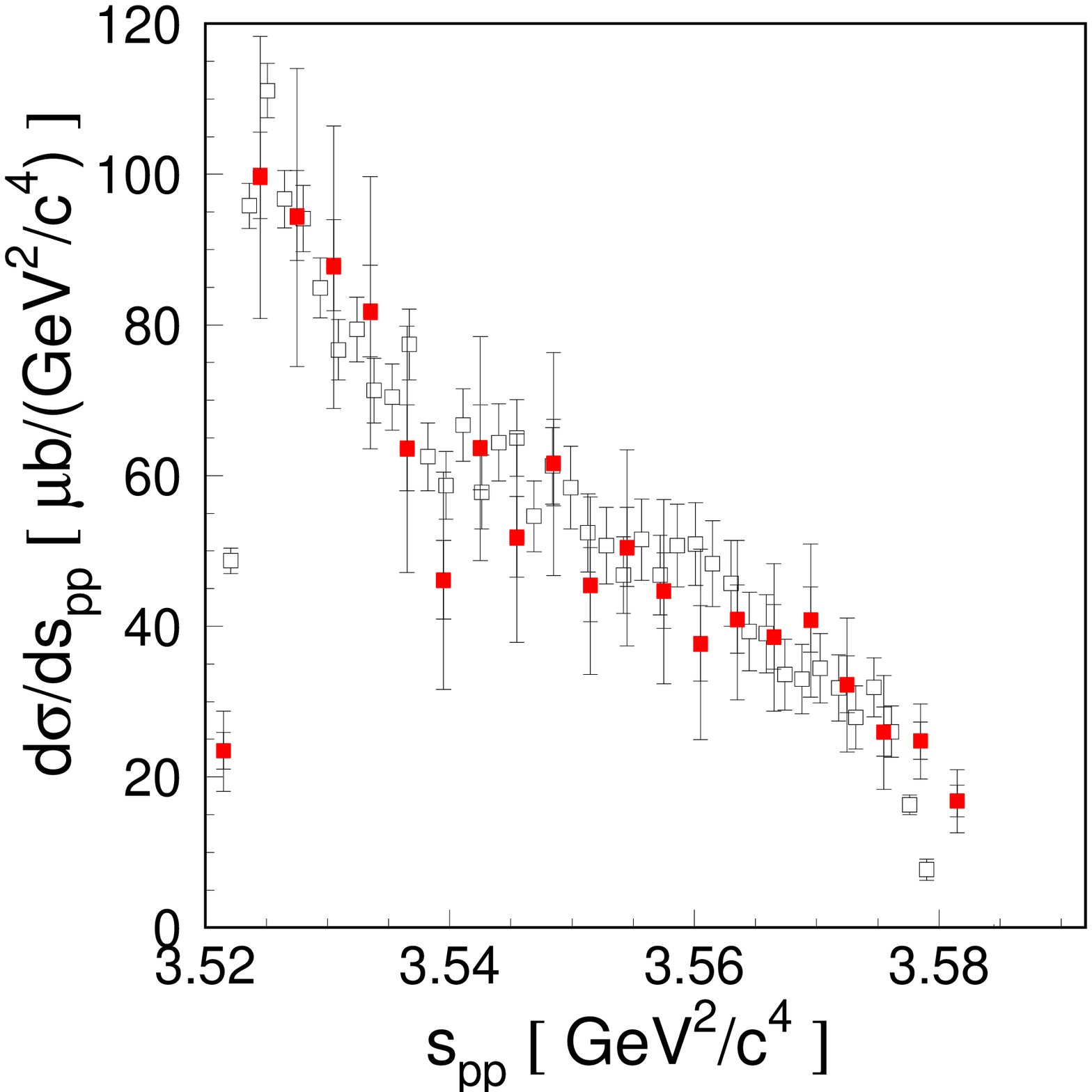} \hspace{-0.9cm}
  \includegraphics[height=.33\textheight]{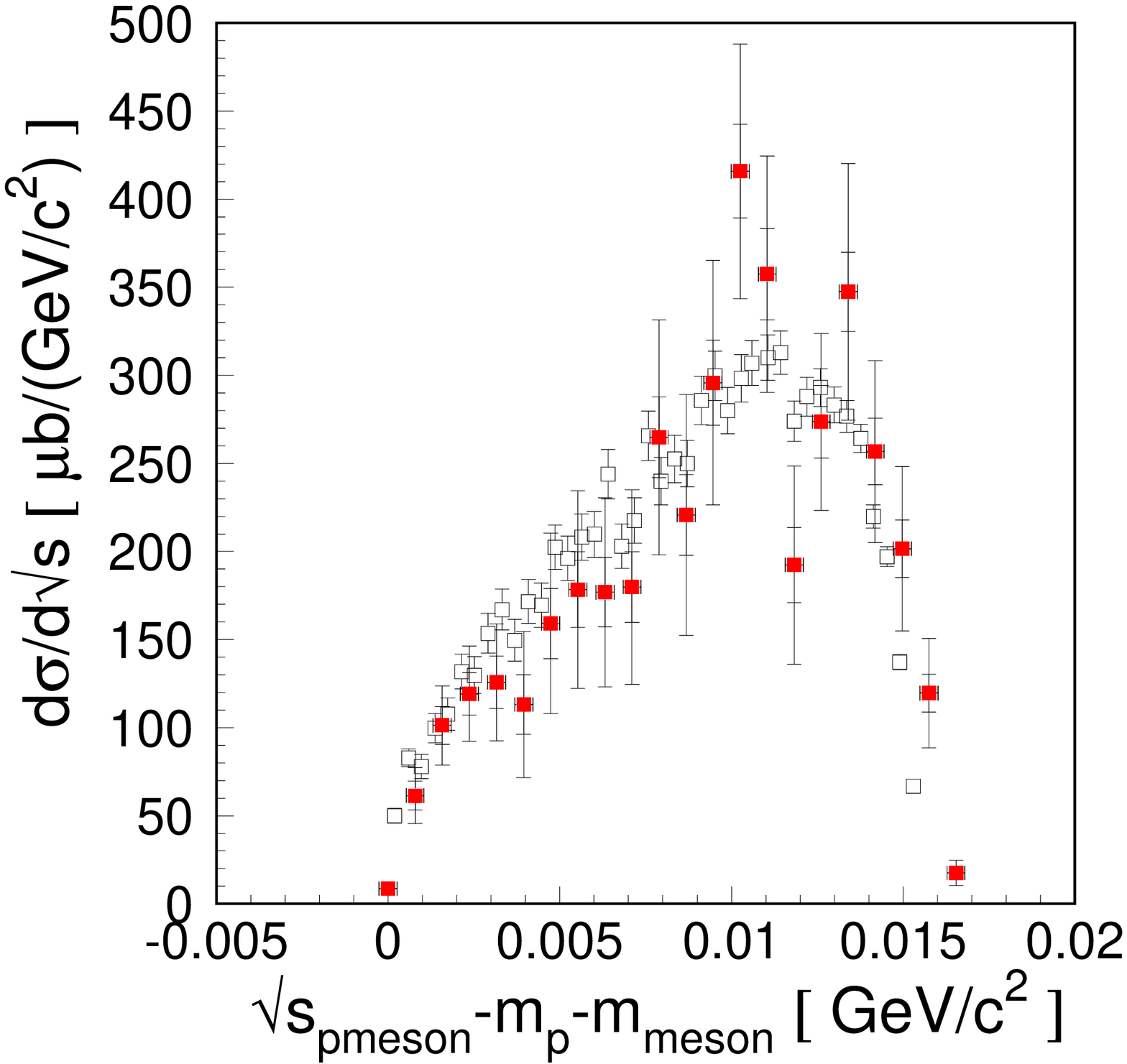}
        \caption{Comparison of the distributions of the squared proton-proton ($s_{pp}$)
        and proton-meson ($\sqrt{s_{p-meson}}$) invariant masses determined
        experimentally for the $pp \to pp\eta$ (full red squares) and
        $pp \to pp\eta'$ (open squares) reactions. The
        distributions for the $pp\to pp\eta^{\prime}$ reaction were normalized in amplitude to the distributions for the  $pp\to pp\eta$ process.}
\label{comp}
\end{figure}
The $pp \to pp\eta'$ reaction has been measured using the COSY-11
detectector setup \cite{brauksiepe, pk, js}. The experiment was based on the measurement
of two protons in the exit channel 
and the 
unobserved meson was identified using the missing mass technique. 
The analysis of the data was described in several
references \cite{menu07,acta,klaja}, and here we would like to present only the
final distributions of the square of the proton-proton ($s_{pp}$)
and proton-meson ($s_{p-meson}$) invariant masses.\\  
In figure \ref{comp} we compare the distributions of the square of the
proton-proton ($s_{pp}$) and proton-meson ($s_{p-meson}$) invariant masses determined
	for the $pp \to pp\eta$ and $pp \to pp\eta'$ reactions. In both panels of the figure, 
	it is seen that the experimental points indicating the $pp \to pp\eta$
measurement are in agreement with those from the $pp \to pp\eta'$ reaction within the
statistical errors.\\
 Unexpectedly, the shapes do not differ, showing enhancements
at the same values of the square of the proton-proton ($s_{pp}$) invariant mass. 
If indeed the $\eta'$-proton interaction is much smaller than the $\eta$-proton
as inferred from the excitation function, then the spectra presented in this report
rather exclude the hypothesis that the enhancement is due to the interaction
of the meson and the proton. 

\section*{Acknowledgments}
We acknowledge the support by the
European Community
under the FP6 programme (Hadron Physics,
RII3-CT-2004-506078), 
by the German Research Foundation (DFG),
and by
the Polish Ministry of Science and Higher Education under grants
No. 3240/H03/2006/31, 1202/DFG/2007/03, 0084/B/H03/2008/34.

\end{document}